\documentclass[11pt]{aa}
\usepackage[english]{babel}
\usepackage{graphicx}

\onecolumn
\begin{document}

\title{Changes of the optical spectrum of the hypergiant $\rho$\,Cas due to a shell ejection in 2013}
\author{V.G.\,Klochkova,   V.E.\,Panchuk, and N.S.\,Tavolzhanskaya}
\institute{Special Astrophysical Observatory RAS, Nizhnij Arkhyz, 369167 Russia}

\date{\today} 

\abstract{Spectral monitoring of the yellow hypergiant $\rho$\,Cas with the by 6-m telescope of the Special
Astrophysical Observatory with a spectral resolution of R$\ge$60\,000 has led to the detection of new features
in the kinematic state of its extended atmosphere following the ejection of matter in 2013. Significant
changes in the profile of the H$\alpha$ line were detected: the line had a doubled core for the first time
in a 2014 spectrum, an inverse P\,Cygni profile on December~13, 2017, and the profile was again doubled 
on August~6, 2017 and  September~5, 2017, but was strongly shifted toward longer wavelengths, indicating 
a rapid infall of matter. 
Splitting of the profiles of strong, low-excitation absorptions into three components was first detected in 2017. 
There is no correlation between the evolution of the profiles of H$\alpha$ and the splitted absorptions. 
Pulsation-like variability with an amplitude of about 10\,km/s is  characteristic only of symmetric weak and 
moderate-intensity absorption lines. Shell emissions of iron-group elements can be identified in the long-wavelength 
part of a spectrum obtained in 2013, whose intensity decreased until they completely disappeared in 2017. 
In the absence of emission in the cores of the H and K~lines of Ca\,II, emissions of shell metals are visible 
in the wings of these lines.
 }

\authorrunning{\it Klochkova et al.}
\titlerunning{\it The optical spectrum of the hypergiant $\rho$\,Cas in 2013--2017}

\maketitle

\hspace{-2cm}
\section{Introduction}

The cool, extremely luminous star $\rho$\,Cas (Sp=G2\,Iae) belongs to a small group of rare yellow
hypergiants. The topical nature of a detailed study of these objects is due to the fact that such massive
stars are the most probable progenitors of Type~II supernovae (SN~II). Moreover, hypergiants are evolutionarily 
related to extremely luminous variables such as $\eta$\,Car. In the Hertzsprung--Russell (hereafter HR) diagram
hypergiants are close to the luminosity limit of the instability strip for A--M stars [1--3]. In addition
to their high luminosities, yellow hypergiants differ from ordinary supergiants in the high rate of their
mass loss via their stellar winds and the presence of gaseous--dusty circumstellar shells. The instability of
hypergiants is manifest through weak brightness variability (with amplitudes $\approx 0.2\div 0.4^m$), usually
referred to as pulsations. The nature of the pulsations of massive stars in the stage of contraction of the
helium core has been studied by Fadeev [4], who concluded that long-period, radial pulsations in $\rho$\,Cas
have a low probability.

Along with the above manifestations of instability, yellow hypergiants also undergo sporadic variations, called 
``shell episodes'', when the star loses mass especially intensively, becoming enveloped in cool, ejected matter 
forming a pseudo-photosphere for several hundred days. Note that the term ``flare'' is used in the Russian 
literature rather than the English ``outburst''; this is not logical, since the brightness of the star 
drops significantly during the ejection of the shell. It would be more correct to use the terms
``eruption'' or ``ejection''. In the case of $\rho$\,Cas, the last event of this kind occurred in 2000--2001, 
when the star lost up to $3\times10 ^{-2} \mathcal{M}_{\sun}$~[5] and its brightness decreased by 1$^m$. 
Despite its significant loss of mass (the mass-loss rate reaches  $10^{-4} \mathcal{M}_{\sun}$/yr), unlike
other yellow hypergiants, $\rho$\,Cas lacks circumstellar dust (see [6] and references therein). No extended
circumstellar structure has been detected, and the star appears point-like even when observed with high
spatial resolution by the Hubble Space Telescope~[7]. At the same time, a thin circumstellar gas envelope is
present, and is fairly structured, as is shown by CO observations~[8, 9].

In the HR diagram, $\rho$\,Cas is located at the boundary of the Yellow Void~[1], separating hypergiants
and luminous blue variables (LBVs) in their quiescent phase. The amplitudes of pulsations of yellow 
hypergiants apparently increases sharply at the boundary of the Yellow Void, increasing the instability 
of their atmospheres and leading to shell ejection~[1]. The results of long-term spectral monitoring~[10--13] of
V1302\,Aql, which is a close relative of $\rho$\,Cas, are relevant here. This yellow supergiant, associated with
the powerful IR source IRC\,+10420, traversed a path in the HR diagram from the region of red supergiants
to the cool boundary of the Yellow Void over the last decade~[13]. 

High-spectral-resolution monitoring of $\rho$\,Cas  has been carried out on the 6-meter telescope of the 
Special Astrophysical Observatory since 2007.  Spectroscopy of $\rho$\,Cas made on this telescope in 
2007--2011 were analyzed and presented in~[14]. Observations over a wide range of wavelengths with
spectral resolution R$\ge$60\,000 have enabled detailed studies of features in the optical spectrum and the
detection of previously unknown properties of the kinematic state of the extended stellar atmosphere.
The heliocentric radial velocity derived from symmetric absorption lines of metals varies with an amplitude 
of about $\pm7$\,km/s relative to the center-of-mass velocity of the system (the systemic velocity) 
Vsys\,=$-47$\,km/s. This time variability is a consequence of low-amplitude pulsations of near-photospheric 
layers of the atmosphere. At some epochs, a velocity gradient is observed in deep layers of the stellar atmosphere. 
Klochkova et al.~[14] found a weak stratification of the velocities in the stellar atmosphere for the first time, 
manifest as a difference of 3--4 km/s in the velocities measured from absorption lines of neutral atoms and ions. 
These authors also showed that the long-wavelength components of the splitted absorptions of Ba\,II, Sr\,II, Ti\,II, 
and other strong lines whose lower levels have low excitation potentials are distorted by a stationary emission  in the
short-wavelength wings of these components.
These non-trivial monitoring results serve as a stimulus for continued spectral monitoring of 
$\rho$\,Cas. In addition, photometric monitoring of the star (AAVSO) indicates that a new ejection of matter
occurred in 2013, during which the brightness of the star decreased by 0.5$^m$. This event occurred only
12 years after the previous one in 2000--2001. Thus, the ejections in $\rho$\,Cas are becoming more frequent.
According to~[5, 15], this may indicate that the star is approaching a crossing of the Yellow Void. This
fact suggests the urgent need of further monitoring of $\rho$\,Cas.

Numerous data have been published for this star, including high-resolution spectral data, however current 
notions about the kinematic state of its extended, unstable atmosphere do not fit any one, generally accepted 
interpretation. Resolution of this question also requires long-term, high-quality observations. We
present here results based on spectral monitoring in 2013--2017. Section~2 briefly describes the methods
used in the observations and analysis of the data. Section~3 presents our results and compare them with
earlier data. Main conclusions are presented in Section~4.

\section{Observations, reduction and analysis of the spectra}

New spectral data for $\rho$\,Cas were obtained at the Nasmyth focus of 
the 6-m telescope of the Special Astrophysical Observatory (SAO) of the Russian Academy 
of Sciences using the echelle spectrograph NES~[16, 17]. The observation dates, 
Julian dates (JD), and recorded spectral ranges are listed in Table~1. Observations were 
made using a large format 2048$\times$4608 E2V CCD42-90-dd CCD. 
To decrease light losses at the entrance slit of the spectrograph without loss of spectral resolution, 
the NES is equipped with a three-slices image scanner. Each spectral order in the image is repeated three
times with a shift along the dispersion of the echelle grating~[17]. The transition to the 
large-format CCD substantially broadened the simultaneously recorded wavelength interval: 
for instance, $\Delta\lambda 5400\div8479$\,\AA{} in the spectrum of October~1,~2014 or 
4697$\div$7782\,\AA{} in spectra obtained in 2017. The spectral resolution is $\lambda/\Delta\lambda\ge$60\,000, 
and the signal to noise ratio is S/N$\ge$100.

The extraction of the one-dimensional spectra from the two dimensional echelle frames was 
carried out using the modified~[18] ECHELLE software of the MIDAS package. 
Tracks of cosmicray particles were removed via median averaging of two successively obtained spectra. 
The wavelength calibration was carried out using the spectrum of a Th--Ar lamp with a hollow cathode. 
Positional and spectrophotometric measurements in the one-dimensional extracted spectra were made using a
modern version of the Dech20t package~[19]. The instrumental match of the stellar spectra and the
calibration spectrum was checked using telluric lines of O$_2$ and H$_2$O. The Vr values were checked using
measurements of up to 70--80 lines in the spectra of $\rho$\,Cas. The rms error in the Vr values derived from
narrow telluric absorption lines is $\le0.5$\,km/s (the accuracy for a single line). The accuracy obtained for
$\rho$\,Cas is somewhat worse due to broadening of the spectral lines by turbulence in the stellar atmosphere.
The procedured used to measure the radial velocities Vr from the NES spectra and the main sources of
errors are described in more detail in~[20, 21].

{
\hspace{-3cm}
\begin{table}[ht!]
\medskip
\caption{Heliocentric radial velocities of $\rho$\,Cas. Vr(sym), Vr(blue), Vr(red) and Vr(emis) 
        are the average velocities derived from symmetric absorption lines, short-wavelength
        components, long-wavelength components of splitted absorption lines, and weak emission lines of metals. 
        The numbers of lines used for the measurements are listed in parentheses.} 
\begin{tabular}{ l c|  c|   c |   c|  l|   c }
\hline
Date &$\Delta\lambda$, &\multicolumn{5}{c}{\small  Vr, km/s} \\ 
\cline{3-7}
JD 2450000+  & \small nm   &   sym  & blue &  red & \hspace{0.5cm} emis & H$\alpha$\\
\hline   
02.02.2013 & 392--698 & $-41.79\pm0.06$\,(494) &$-60.9\pm0.4$\,(16) &$-34.8\pm 0.3$\,(22)&  $-49.8\pm 0.3$\,(3) &$-40.1$   \\ 
6326.29  &&&&&& \\
\hline 
08.08.2014 &392--698 & ${\bf -42.37\pm0.12}$\,(226),& $-79.0\pm0.2$\,(49), &$-33.2\pm 0.2$\,(43)& $-50.2\pm0.7$\,(3)& $-60.3$ \\ 
6877.53&             &$-32.18\pm0.14$ (249)         &   $-59.4\pm0.4$\,(20)         && &\\ 
\hline
01.10.2014 &540--848 & $-36.35\pm0.14$\,(181)  &$-81.0\pm0.5$\,(6)          & $-35.8\pm0.4$\,(12)& $-51.9\pm 0.5$\,(11)&$-59.3$  \\ 
6931.57 &&&&&& \\
 \hline
13.02.2017& 470--778 & $-39.61\pm0.06$\,(447)  & $-67.7\pm0.3\,(13)$, & $-33.7\pm0.3$\,(13)   & $-52.2\pm 0.2$\,(7)& $-41.3$ \\ 
7797.59&          &                         & $-55.3\pm0.1$\,(13) &                       && \\ 
\hline                                                                                                                            
06.08.2017& 470--778 &  $-50.79\pm 0.09$\,(374)&$-66.7\pm0.2\,(18)$,& $-31.2\pm0.3$\,(14)& $-48.2\pm0.6$\,(6) & $-31.4$ \\ 
7971.50 &          &                           & $-56.4\pm0.3$\,(13)&                   &  & \\
\hline
06.08.2017& 470--778 & $-50.60\pm 0.07$\,(396) &$-67.1\pm0.2$\,(25),& $-31.0\pm0.3$\,(20)  & $-46.8\pm0.1$\,(5) & $-31.2$ \\ 
7972.34 &          &                         & $-56.3\pm0.2$\,(20) &                      & &\\ 
\hline
13.08.2017& 470--778 & $-49.51\pm 0.07$\,(349)& $-66.9\pm0.2\,(21)$, & $-32.1\pm0.3$\,(22)  & $-50.1\pm1.0$\,(4) & $-31.8$ \\ 
7979.30&          &                        & $-56.5\pm0.3$\,(12)  &                      &  &\\ 
\hline
05.09.2017& 470--778 & $-47.85 \pm 0.07$\,(355)& $-67.0\pm0.2\,(28)$, & $-33.0\pm0.3$\,(16) & &$-46.1$  \\ 
8013.52 &          &                        & $-56.3\pm0.2$ (19)   & & \\               
\hline
\multicolumn{7}{l}{\it\footnotesize  The value derived from the lines of ions on August 8, 2014 is shown in boldface. }\\ 
\end{tabular}   
\label{velocity}
\end{table}      
}

\section{Discussion of results}  

When analyzing the observations, we measured radial velocities using different types lines in the spectrum 
of $\rho$\,Cas selected in accordance with [14]: weak symmetric absorptions, components of splitted 
absorption lines of metals, and the H$\alpha$ line with its variable absorption-emission profile. 
In addition to the these types of lines, the radial velocities were also measured using the weak, narrow emission
lines of metals listed in Table~2. All these types of spectral fratures were already described in the earliest
studies of the spectrum of this star~[8, 22]. Let us consider in more detail the properties of the profiles of
the above types of lines and the measurement results.

\subsection{Metallic emissions}

The wide range of wavelengths recorded in the spectra of $\rho$\,Cas obtained on October~1, 2014 and
February~13, 2017 enabled measurement of the positions of several reliably selected weak emission
lines of Fe, Co, and Ni with low excitation potentials $\chi_{low} \le 3$\,eV (see Table~2). 
The average radial velocity derived from these emission lines, presented in Table~1, 
varies only slightly with time, and differs little from the systemic radial velocity
Vsys\,=$-47$\,km/s [24]. The metal emission lines have widths of about 10\,km/s, 
appreciably exceeding the width of the spectrograph instrumental profile 
$\delta Vr \approx 5$\,km/s. The small widths of these emission lines and the coincidence of V(emis) 
with the systemic velocity indicates that these lines are formed in an external,
extended gas envelope, whose dimensions substantially exceed the photometric radius of the star. 
If this envelope is optically thin in these emission lines, their positions indicate the 
systemic velocity of the star. This value remains constant within our uncertainties,
and the rate of expansion of the shell is several km/s. Emission lines are observed predominantly 
during periods when the stellar brightness is decreasing; this may indicate relative stability 
of the emission measure, observed against the background of the weakened photospheric spectrum. 
Measurements of the intensity of emission lines relative to the local continuum are not sufficient 
to draw conclusions about their variability, since the level of the continuum can vary.

\begin{figure}[ht!]
\includegraphics[angle=0,width=0.9\textwidth,bb=40 45 730 530,clip]{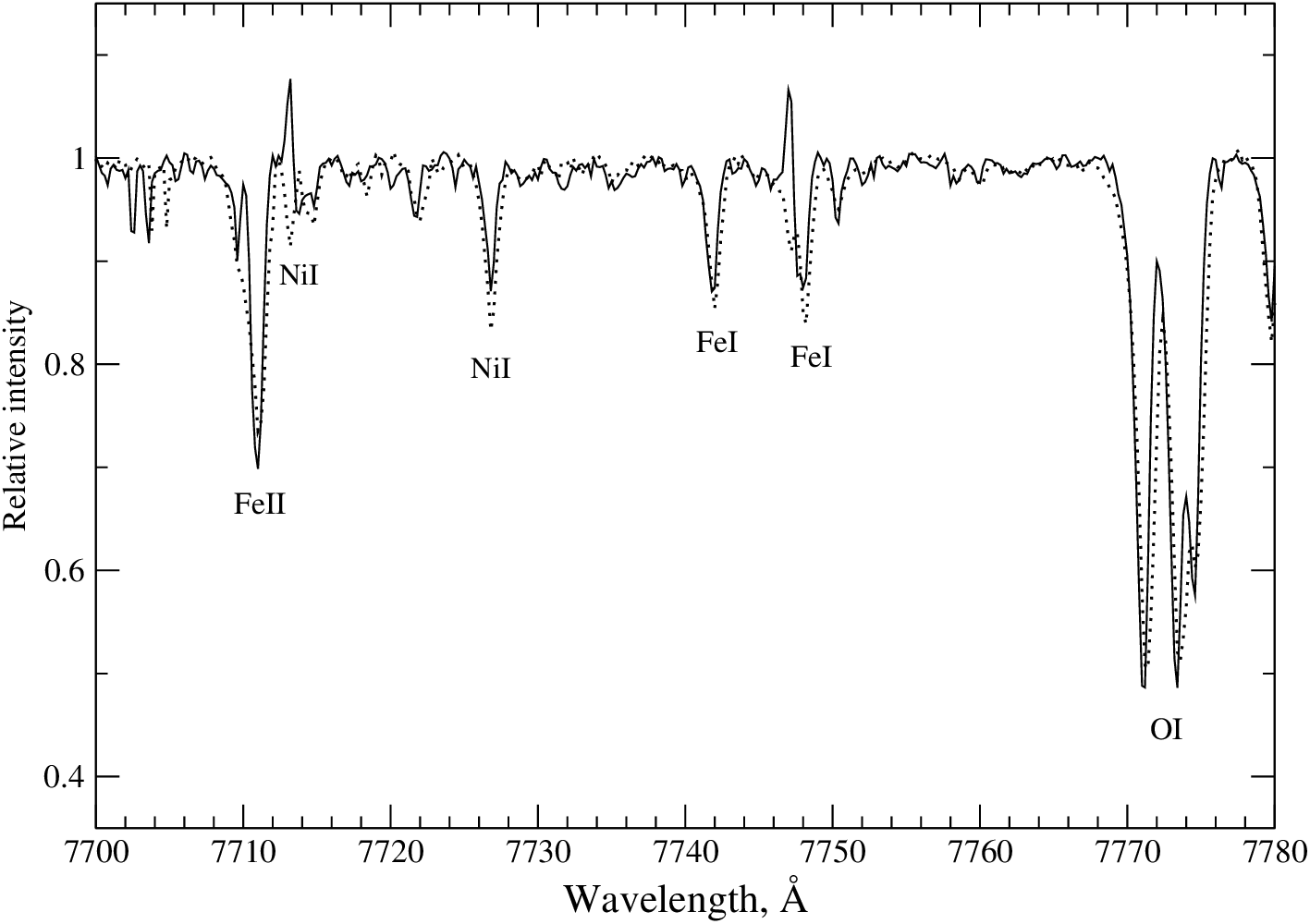}   
\caption{Fragments of the $\rho$\,Cas spectra at wavelengths 7696--7773\,\AA{}, 
   obtained on February~13, 2017 (solid) and August~6, 2017 (dotted). The main 
   identified absorption lines are marked. The fragment of the February~13, 2017 spectrum 
   contains two weak emissions of NiI and FeI  } 
\label{7700}   
\end{figure}

\begin{table}[ht!]
\medskip
\caption{Emission lines of metals in spectra of ρ Cas obtained at different epochs. Excitation potentials of 
lines  $\chi_{\rm low}$    were taken from the VALD database~[23]. Conﬁdently identiﬁed emission lines are 
    marked by a ``+'', and uncertain features by a ``+:''. A sign “$-$” indicates the absence of a given feature. } 
\begin{tabular}{c  c  c   c  c l c  c}
\hline
Line   & $\chi_{\rm low}$ & 1.10.2014& 13.02.2017 &06.08.2017 & 13.08.2017&05.09.2017 \\  
\hline
 FeI 8047.617   &  0.85  &    +&       &    &   &\\  
 FeI 7912.866   &  0.86  &    +&    +  &    &   &\\  
 FeI 7748.269   &  2.94  &    +&    +  & $-$&$-$&$-$\\  
 NiI 7714.308   &  1.94  &    +&    +  & $-$&$-$&$-$ \\ 
 FeI 7583.788   &  3.02  &    +&    +  &  + & + &$-$\\  
 NiI 7291.449   &  1.94  &    +&    +  &  + & + &$-$ \\ 
 CoI 7052.870   &  1.96  & $-$ &    +  &  +:& +:&$-$ \\ 
 CoI 6814.950   &  1.96  &   +:&    +  &  + & + & +: \\ 
 FeI 6574.240   &  0.99  & $-$ &    +  &  + & + & +: \\ 
 FeI 6498.950   &  0.96  & $-$ &    +  &  + &$-$& $-$ \\ 
 FeI 6358.693   &  0.85  &  +  &    +  &  + & + & + \\   
 NiI 6108.107   &  1.67  & $-$ &    +  & $-$&$-$& $-$ \\ 
 FeI 5956.700   &  0.86  & $-$ &    +  &  + & + & + \\   
\hline                             
\end{tabular}                                 
\label{emis} 
\end{table}

As follows from Table~2, shell emissions of metals are most numerous in spectra obtained 
after the deep brightness minimum in 2014. After the brightness maximum, closer to Autumn 2017, 
the relative intensity of the metal emissions has decreased, and some have become completely absent 
from the spectra. All shell emissions were also absent in our spectra obtained in 2007--2011. 
The variability of the emissions is illustrated in Fig.\,1, which shows a fragment of the spectrum 
of $\rho$\,Cas for two observing dates. The time behavior of the emission lines is consistent with 
the previous description of Scheffer and Lambert~[25]. These authors, who obtained a large collection 
of spectra of $\rho$\,Cas, estimated the dominant period of stellar pulsations to be $\approx 500^d$. 

Emissions of this type were also detected in the spectrum of $\rho$\,Cas earlier
by Sargent~[22] and Gorlova~et\,al.~[9]. All these authors also observed variability of weak 
emissions. Gorlova~et\,al.~[9] noted the appearance of emissions just after the 
brightness maximum, after which they weakened and disappeared as the brightness decreased. 
We have found a similar behavior of the emission lines in relation to the variability of the 
stellar brightness; as follows from the AAVSO database (www.AAVSO.org), both of our 2014~spectra 
and the spectrum for February~2017 were obtained when the brightness of $\rho$\,Cas was increasing. 
The spectra obtained in August--September~2017 correspond to the period of brightness decrease, 
when the intensity of the metallic emission lines was decreasing. 
A comparison of the data in Tables~2 and~1 shows that, at times when the largest numbers 
of emissions were detected (October~1, 2014 and February~2, 2017), the atmosphere was in 
the phase of compression: the radial velocity Vr(sym) differed by more than +10\,km/s 
from the systemic velocity. Since August~2017, velocity Vr(sym) has been close to 
the systemic velocity and the intensity of the emissions has been decreasing, sometimes 
leading to their disappearance. It is obvious that the presence or absence of weak metal 
emissions depends on the flux level in the continuum. At times close to the maximum
brightness, emissions are lost against the high flux background. The emissions 
become most intense during periods of powerful ejections of matter, which lead to weakening 
of the star brightness.

According to~[26], the H$\alpha$ emission lines forms in the outer layers of the atmosphere, 
which are thermally excited by shocks. However, the observation of chromospheric effects 
in a star with such a vigorous optically dense pseudo-photosphere as that of $\rho$\,Cas
is unlikely. We attempted to search for traces of chromospheric emission in the core of 
the K line in the spectra taken on February~2, 2013 and August~8, 2014, where the recorded 
wavelength range includes the CaII resonant doublet. We confidently detected
the remnants of the chromospheric emission profile for the low-luminosity objects, 
even at a level of 0.04 of the local continuum (see, e.g., Fig.\,1a in~[27]).
However, in the spectrum of $\rho$\,Cas, only emissions of other metals with radial 
velocities close to the systemic velocity determined from the emission lines
listed in Table~2 are observed against the background of the CaII line profile.

Since the stellar radiation in the cores of the resonance lines is more than 4$^m$ weaker 
than the local continuum, emission lines of metals close to the cores of the CaII~H and K~lines 
are visible independent of the brightness variations of $\rho$\,Cas. For instance, no
emission lines were present in the red part of spectrum of $\rho$\,Cas obtained on February~2, 
2013, just before the ejection episode. However, on both this date and  on August~8, 2014, 
we detected the CaII~H and K~lines in the short-wavelength part of the spectrum, in whose 
wings narrow metal emission lines originating in the shell are clearly visible,  similar 
to those detected in the red part of the spectrum. The 6-th column of Table~1 lists the 
mean velocities V(emis) derived from the emission lines of FeI in the wings of the CaII~H
and K~lines: 3927.92, 3930.30, and 3969.26\,\AA{}.

We emphasize that there  is no emission in the cores of the CaII~H and K~lines, indicating the absence of 
a chromosphere, as it was noted earlier by Lobel~et\,al.~[5].

\begin{figure}[h!]
\includegraphics[angle=0,width=0.5\textwidth,height=0.4\textwidth,bb=40 50 710 550,clip]{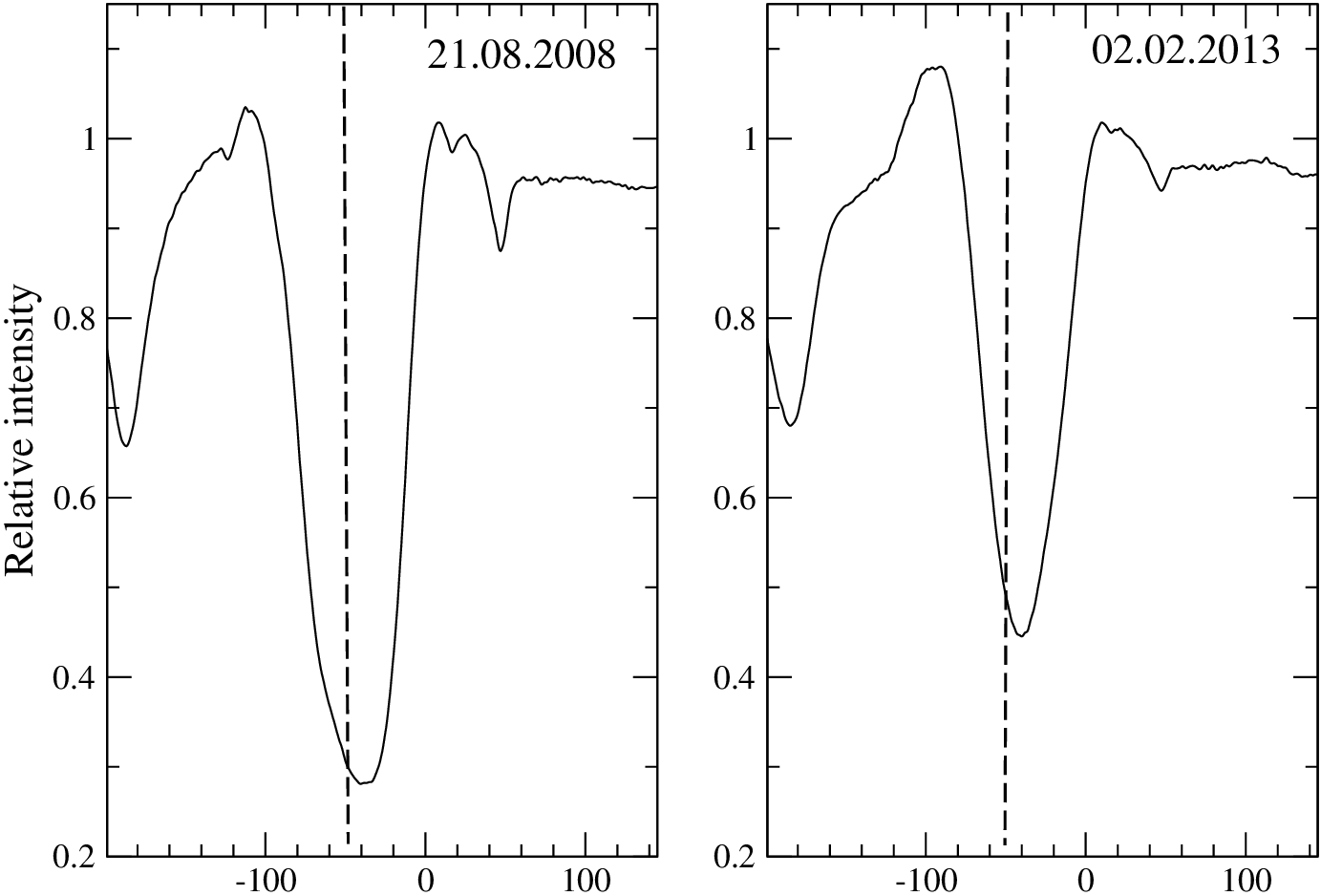} 

\includegraphics[angle=0,width=0.5\textwidth,height=0.4\textwidth,bb=40 50 710 550,clip]{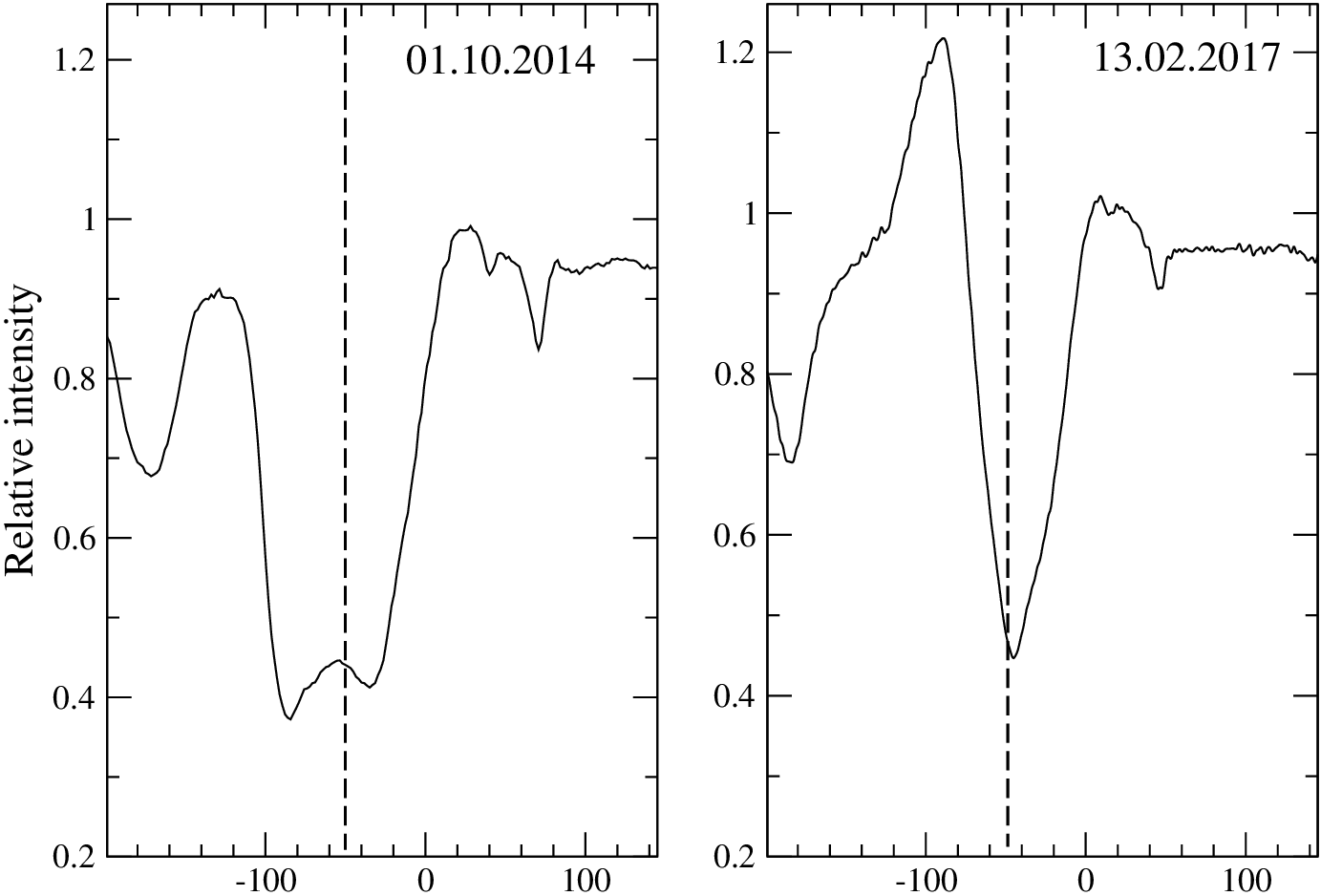} 

\includegraphics[angle=0,width=0.5\textwidth,height=0.4\textwidth,bb=40 50 710 550,clip]{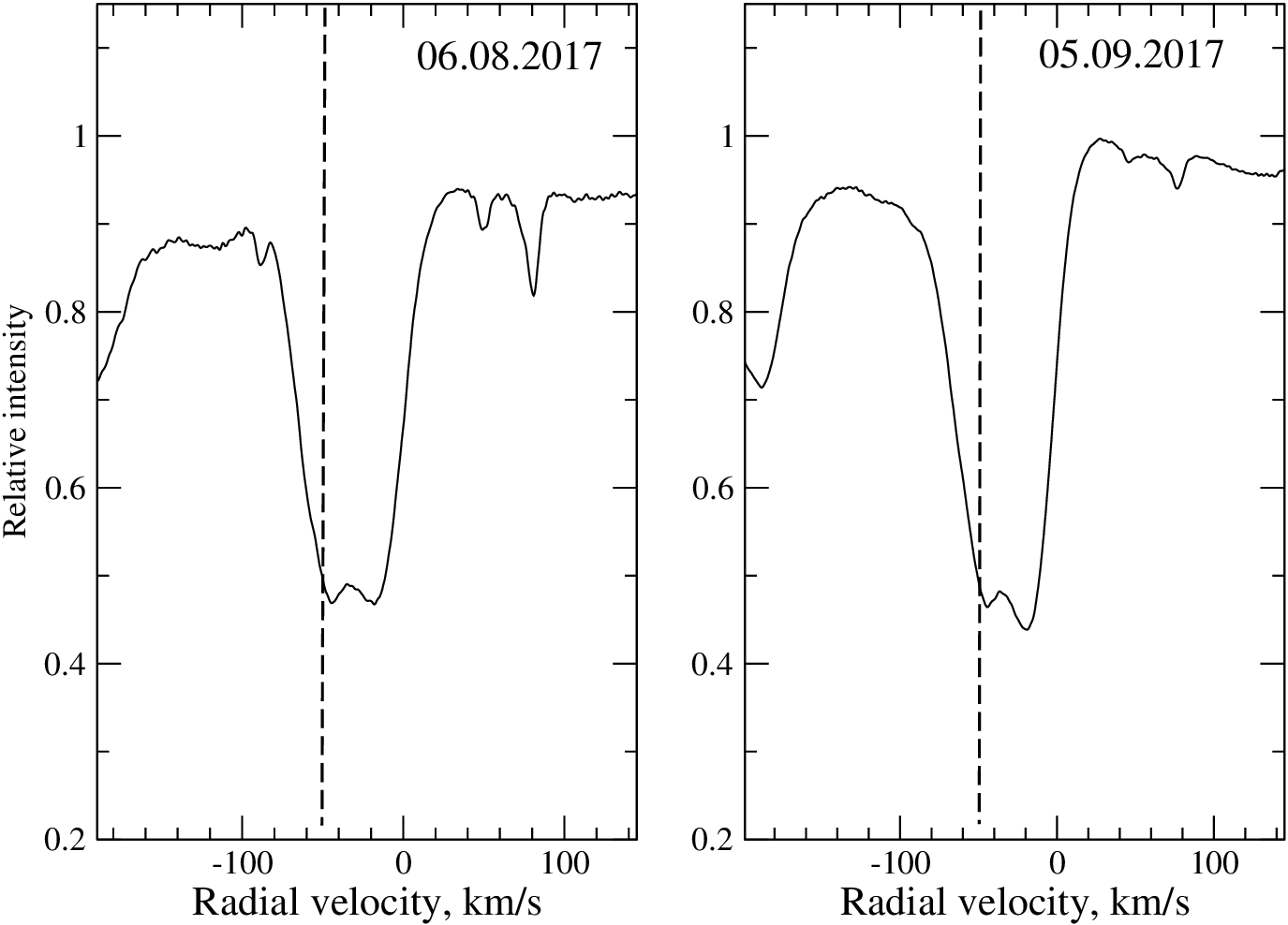} 
\caption{H$\alpha$ profile in spectra of $\rho$\,Cas obtained on different dates. The vertical dashed 
   line shows the systemic velocity Vsys\,=$-47$\,km/s~[24].} 
\label{Halpha3}   
\end{figure}

\subsection{Behavior of the H$\alpha$ profile}  

For all observing dates in 2013--2017, the spectral interval recorded contains the H$\alpha$ line, 
whose complex absorption--emission profile is formed in high layers of the extended shell, 
which are affected by the stellar wind and shocks. As can be seen in Fig.\,2, the position 
of the absorption core, the intensity of the emission components, and the ratio of their 
intensities vary with time. The variations of the position of the H$\alpha$ absorption core 
indicates that the region in which it is predominantly formed in the stellar atmosphere shifts. 
The type of the H$\alpha$ profile changed appreciably in the course of the new set of observations, 
as is clearly shown in Fig.\,2. 

In 2008--2011, the H$\alpha$ profile represented almost symmetric absorption with its 
core close to the systemic velocity, as in the H$\alpha$ profile in the spectrum taken on 
August~21, 2008 (Fig.\,2). However, in the spectrum taken on February~2, 2013, even before 
the beginning of the shell episode, the H$\alpha$ profile shifted toward longer
wavelengths, indicating the onset of the collapse of the atmospheric layers in 
which this line forms. Just after the shell episode, on October~1, 2014, the profile
again changed significantly: the line has a doubled core whose short-wavelength component is
shifted by up to 50\,km/s respective to the systemic velocity. In 2017, the H$\alpha$ 
line resembled an inverse P\,Cygni profile (February~13, 2017). It then again became doubled, 
with a large shift, but this time toward longer wavelengths (August~6, 2017). This testifies 
to a contraction of the upper layers of the atmosphere. This behavior of the H$\alpha$ 
profile indicates increased instability of the upper layers of the stellar atmosphere 
before the ejection and during the restoration of the brightness following this episode.

Dates of observations when significant H$\alpha$ emission appears are of the most interest. 
In our new set of spectra, emission (at $\approx20$\% of the continuum level)
was detected on February~2, 2013 and on February~13, 2017. Both profiles are shown in Fig.\,2. 
As follows from the AAVSO data, both observing moments correspond to the ascending branch, 
before the maximum brightness.  One of the spectra of our earlier set (years 2007--2011) is 
of particular interest. As can be seen in Fig.\,1 of~[14], in the spectrum for October~20, 2008,
emission features were detected in both wings of the H$\alpha$ line. This epoch corresponds to 
the descending branch of the light curve.

The appearance of H$\alpha$ emission is not related to the properties of the metal-line 
profiles on these dates. As can be seen from the data in Table~1, the profiles of
strongest absorptions have two components in the spectrum for February~2, 2013, while 
they have three components in the spectrum for February~13, 2017 and subsequent dates. 
On the other hand, the H$\alpha$ line has different profiles on February~13, 2017 and
August~6, 2017, while the velocity patterns and strongest absorption profiles for 
these two spectra coincide. The velocities derived from weak symmetric absorption 
lines also differ on these dates. Thus, we have found new evidencce for an absence of a 
direct link between the layers in which H$\alpha$ and strongest absorptions are formed and 
deeper layers that are subject to pulsations.

\subsection{Spliited sbsorptions}

In the spectra taken on all dates in 2014 and 2017, the profiles of strong absorptions in the spectrum
of $\rho$\,Cas remained splitted. A list of spltted absorptions in the spectra of $\rho$\,Cas was
presented earlier in~[14]. The short-wavelength components of the splitted absorptions are formed in the 
uppermost layers of the expanding atmosphere, close to the circumstellar gaseous envelope, as is 
confirmed by the coincidence of the velocities of these components and those derived from CO~lines~[8]. 
However, unlike earlier observing seasons (2007--2011), the strongest absorptions in the 2017 spectra 
have a more complex structure, since the short-wavelength component also displays two
absorption features. This is clearly visible in Fig.\,3, where the profile of the strong BaII~6141\,\AA{} 
absorption line observed on February 13, 2017 is compared to profiles of this line from earler 
observing seasons (2008 and 2013) in a plot of radial velocity versus relative intensity. 
It is necessary to emphasize that the long-wavelength component displays a constant shift relative 
to the systemic velocity by approximately 15\,km/s.

\begin{figure}[ht!]
\includegraphics[angle=0,width=0.7\textwidth,bb=40 50 710 530,clip]{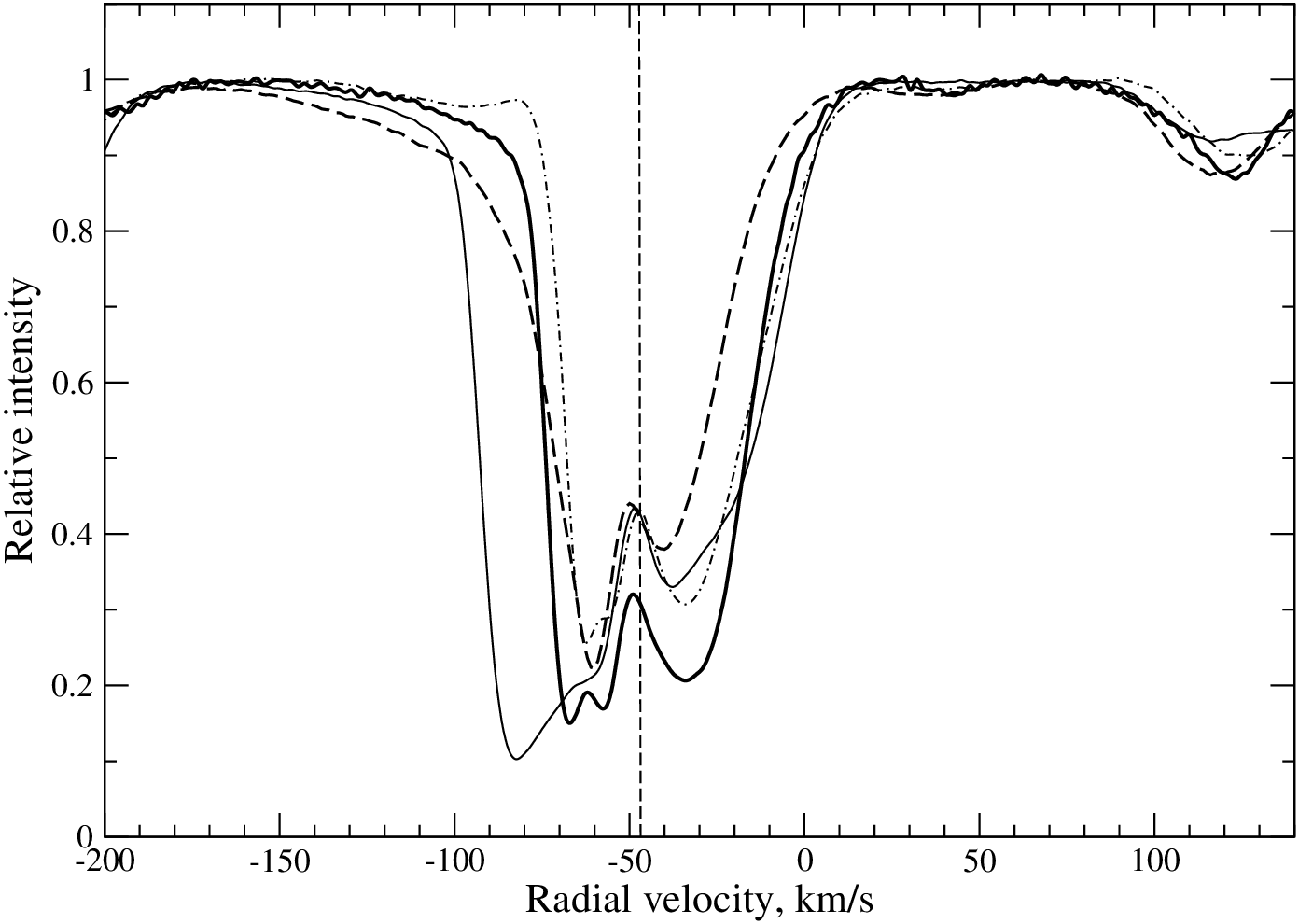} 
\caption{Profiles of the BaII~6141\,\AA{} line in the spectra of $\rho$\,Cas for September~30, 2009 (dash-dot), 
    February~2, 2013 (dashed), August~8, 2014 (thin solid), and February~13, 2017 (bold). The vertical dashed 
    line shows the systemic velocity Vsys\,=$-47$\,km/s~[24].} 
\label{Ba6141}   
\end{figure}

The kinematic instability of the upper atmospheric layers subject to the influence of the stellar wind is
also manifest as variability of the short-wavelength wing of the BaII~6141\,\AA{} line, which indicates 
appreciable strengthening of the wind before the shell ejection, as can be seen in the spectrum taken on
February~2, 2013 in Fig.\,3, even before the shell episode extends to $-150$\,km/s. In the spectrum for
August~8, 2014, during the exit from the minimum state, the shape of the BaII~6141\,\AA{} profile differs
significantly from its shape at the other epochs shown in Fig\,3. The line is still splitted, and the
position and intensity of the long-wavelength component does not differ from their values in the profile
at earlier times. The short-wavelenth component is considerably broadened (by approximately a factor of
two), and is also shifted toward shorter wavelengths, as can also be seen from the data presented in Table~1.

The BaII~6141\,\AA{} profiles in Fig.\,3 illustrate well the difference in the steepness of the wings of the
two components of the splitted absorption, indicating different profile widths and differences in the physical 
conditions in the regions of formation of the components. This confirms that the short-wavelength component 
with its steeper short-wavelength wing is formed in uppermost layers of the extended stratified atmosphere of 
the hypergiant. Additional splitting of the short-wavelength component indicates intensified stratification of 
the uppermost layers of the atmosphere.

\subsection{New traits of the kinematic state of the atmosphere of $\rho$\,Cas}

All our radial-velocity measurements using lines of different types are presented in Table~1 and in
several ﬁgures showing profiles of various spectral features. The amplitude of the velocity variability
we have derived from symmetric absorption lines, Vr(sym), exceeds 10 km/s, which is higher than the
variability amplitude found in our previous study of $\rho$\,Cas. The Vr values for the short-wavelength 
components Vr(blue1) and Vr(blue2) presented in Table~1 differ by more than 12\,km/s. As follows 
from Fig.\,3, this splitting of the line, which was absent from all previous spectra~[14, Fig.\,1], 
is already appreciable in the spectrum taken on September~30, 2009.

\begin{figure} 
\includegraphics[angle=0,width=0.5\textwidth,height=0.3\textheight,bb=40 50 720 530,clip]{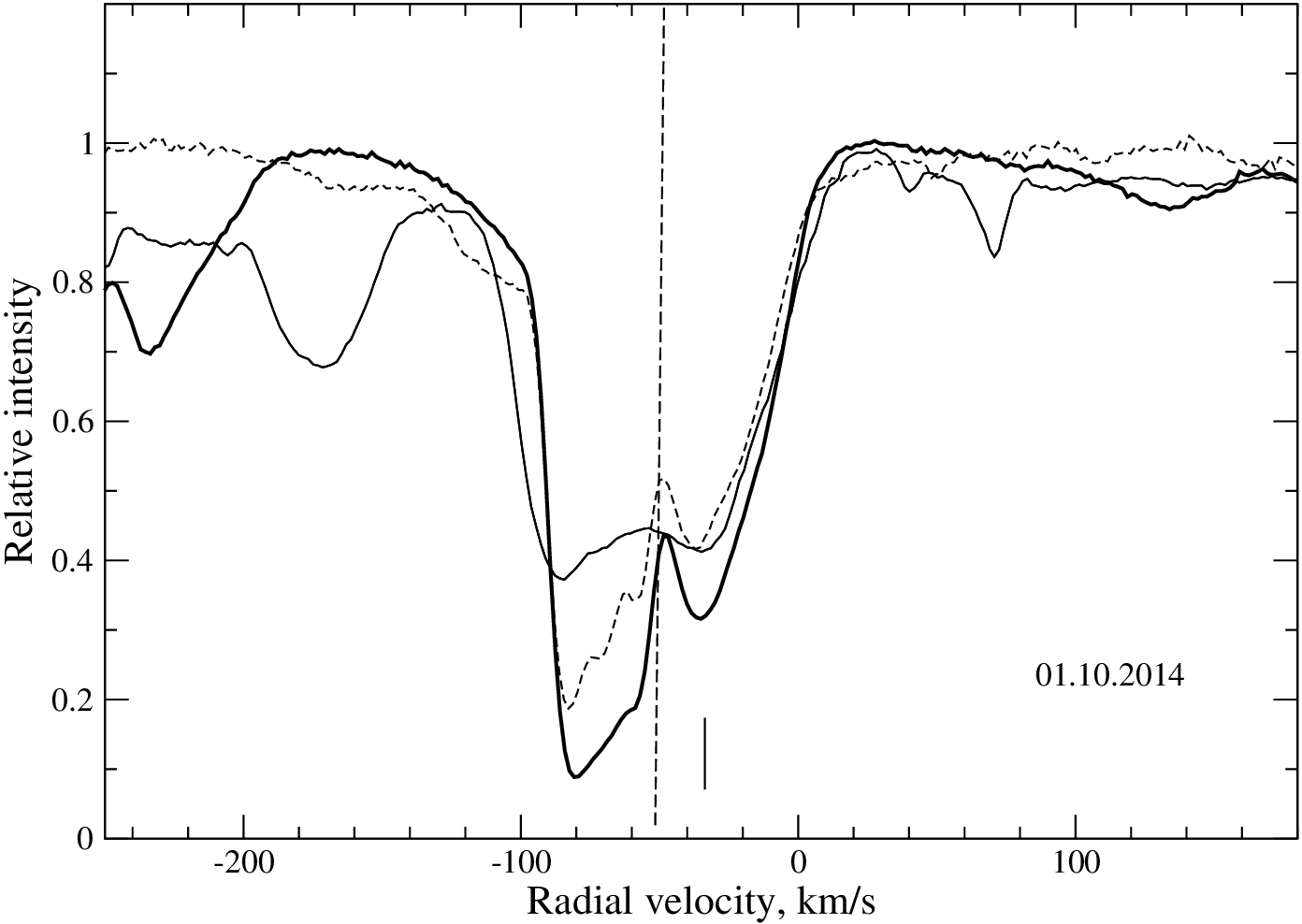}  
\includegraphics[angle=0,width=0.5\textwidth,height=0.3\textheight,bb=40 50 720 530,clip]{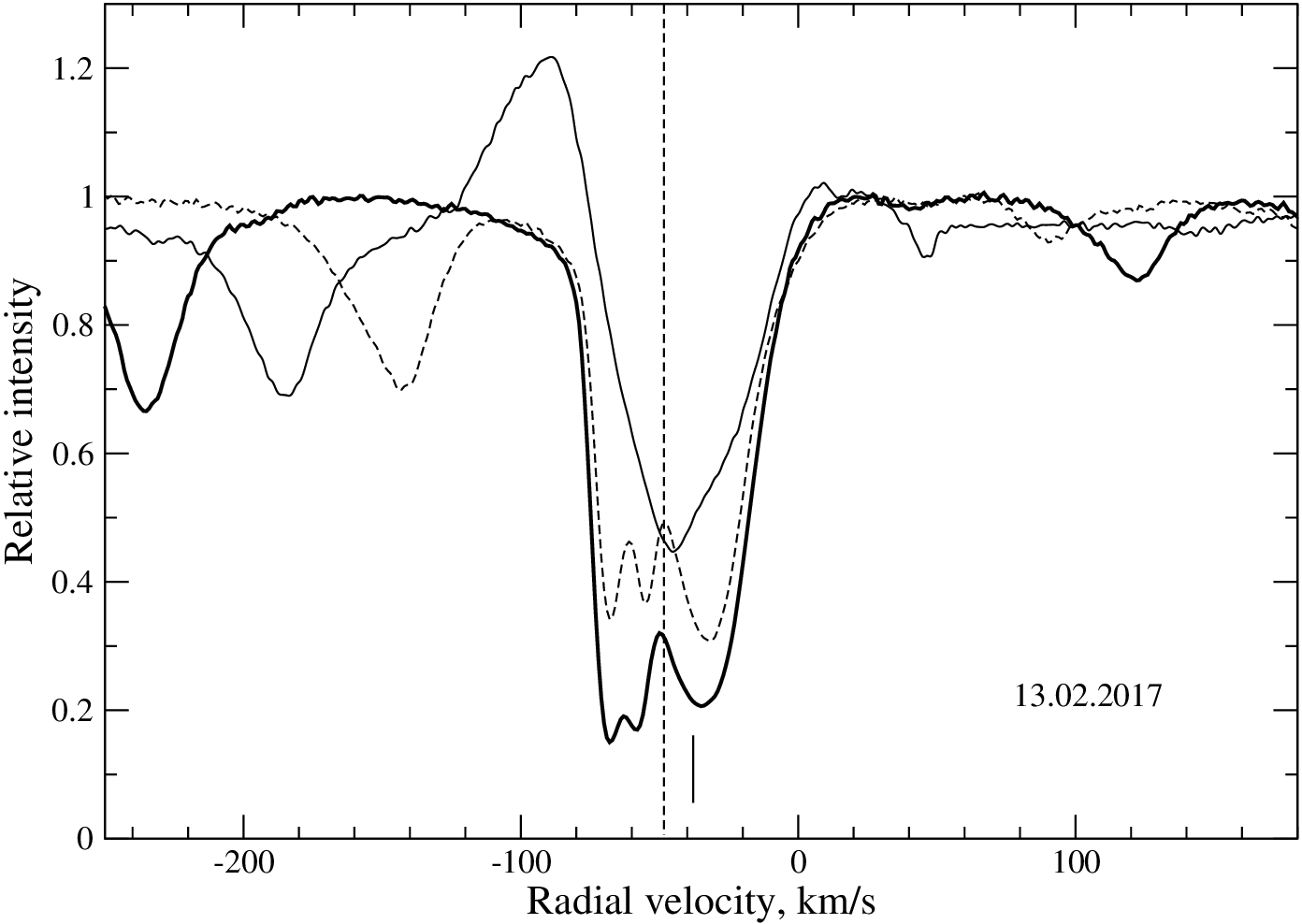}  

\includegraphics[angle=0,width=0.5\textwidth,height=0.3\textheight,bb=40 50 720 530,clip]{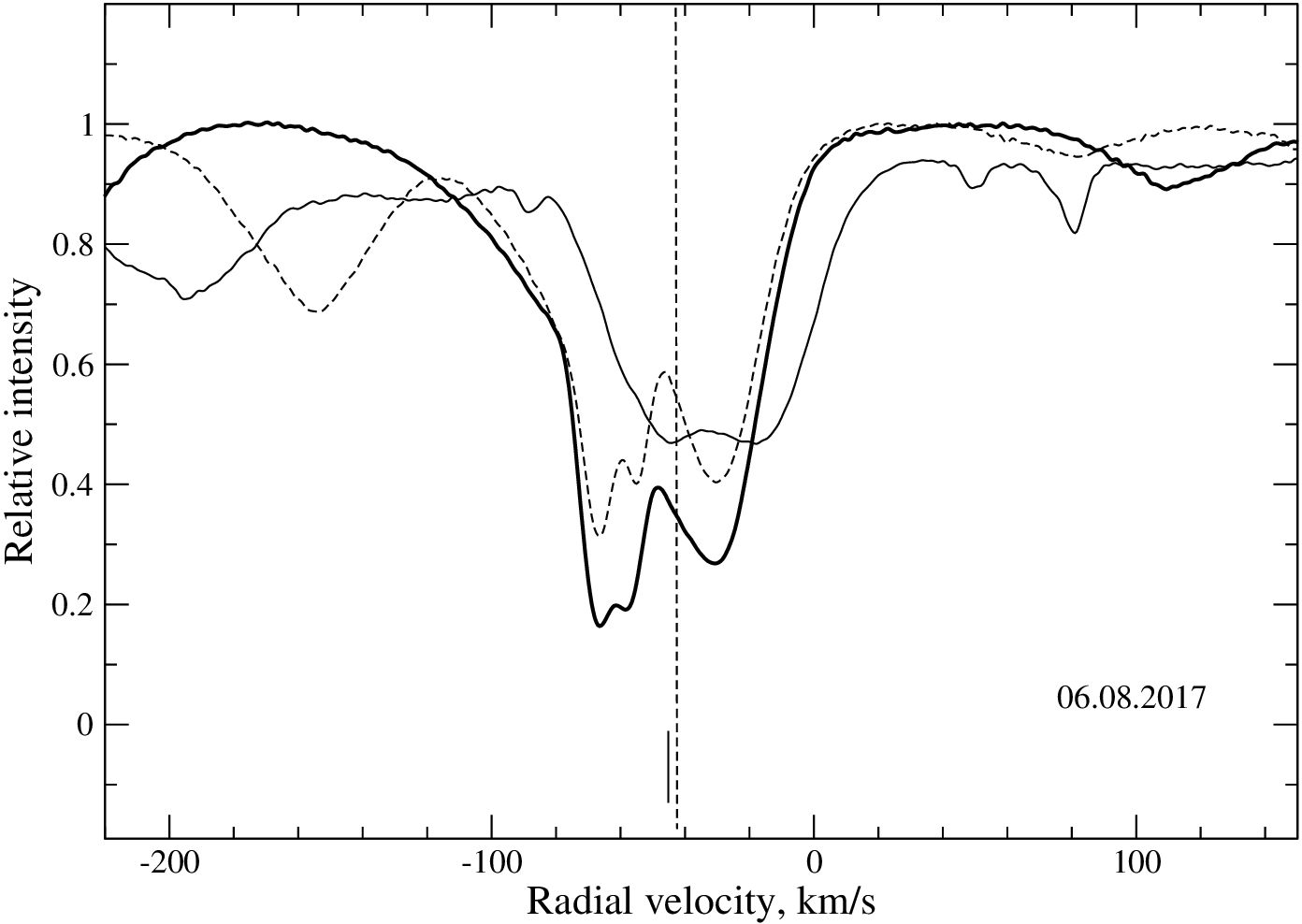} 
\includegraphics[angle=0,width=0.5\textwidth,height=0.3\textheight,bb=40 50 720 530,clip]{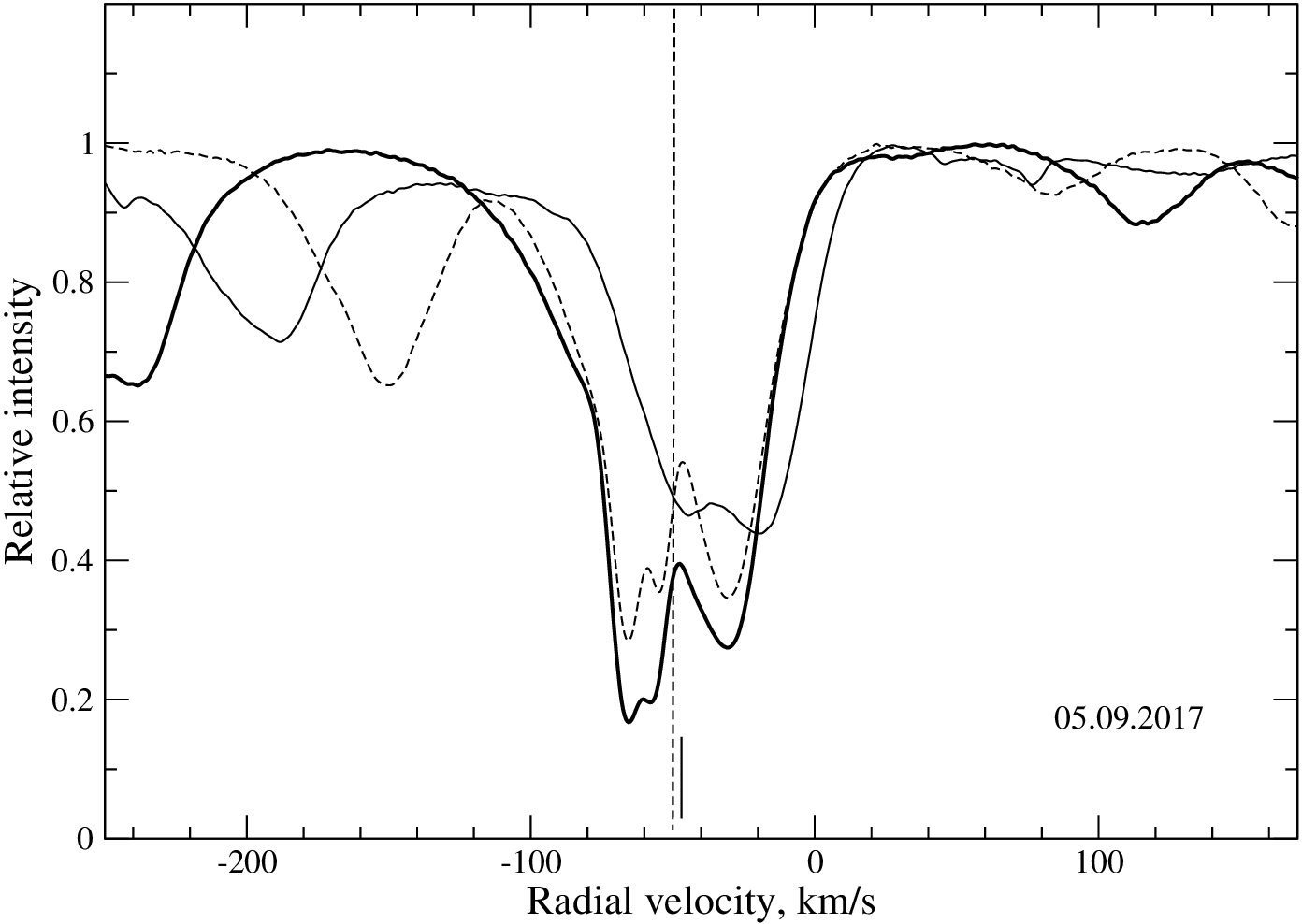}   
\caption{Diversity of the profiles of selected lines in the spectra of $\rho$\,Cas: H$\alpha$ (thin), FeI~5446 (dashed), 
        and BaII~6141 (bold). The vertical dashed line shows the systemic velocity Vsys\,=$-47$\,km/s~[24]. 
        In all panels, the vertical bar indicates the velocity derived from symmetric absorption lines (see Table~1).} 
\label{Profiles}   
\end{figure}

Figure~4 can be used to compare the behavior of the profiles of H$\alpha$ and the splitted absorptions. 
In the 2014 spectrum, before the appearance of H$\alpha$ emission, the position and width of the profile of
this line almost coincided with those for the splitted absorptions. At the epoch when the emission appears 
(February~13, 2017), the shift of the shortest-wavelength component of the splitted absorptions toward 
shorter wavelengths decreases to Vr(blue)$\approx -67$\,km/s. As can be seen from the fourth column of 
Table~1, on subsequent dates, this short-wavelength component and the entire BaII~6141\,\AA{} profile 
remain at the same position, independent of the continuing variations of the H$\alpha$
profile. In general, we conclude that there is no strict correlation between the evolution of 
the H$\alpha$ and BaII~6141\,\AA{} profiles. For all the dates in Fig.\,4, the vertical bar 
indicates the value of Vr(sym) derived from symmetric absorptions. Figure~4 shows an absence of any 
relationship between the position of the H$\alpha$ line and Vr(sym); i.e., pulsations in deep layers
of the atmosphere do not affect the H$\alpha$ profile, which forms in high atmospheric layers.

It is surprising that the positions of strong absorptions were continuously stable. This is clearly
visible in the BaII~6141 and FeI~5446\,\AA{} profiles in Fig.\,4, and the positions of both the long- 
and short-wavelength components for a large set of splitted absorptions in Table~1. This result is also
confirmed by the positions of the long- and short-wavelength components in 2007--2011~[14]. 
However, the positions of the short-wavelength components of the splitted absorptions is subject 
to the influence of the stellar wind, manifest as a distortion of the short-wavelength wings of these lines. 
In essence, we can conclude that the strongest atmospheric absorptions are not influenced by pulsations.

A special velocity pattern was obtained in the spectrum taken on August~8, 2014, at the maximum
brightness of the star. Measurements of a large set of symmetric absorptions in this spectrum 
demonstrate a systematic difference between the velocities derived from absorption lines of ions and neutral 
atoms, by about 10\,km/s. This kind of stratification was found by us earlier for several 
observations of $\rho$\,Cas in 2007--2011 (see~[14, Fig.\,2]). However, this velocity difference 
is particularly large in the spectrum for August 8, 2014: Vr(sym)\,=$-42.37$\,km/s for ions,
while Vr(sym)\,=$-32.18$\,km/s for neutral atoms.
The difference of the radial velocities derived from symmetric absorptions of ions and atoms is also
visible in Fig.\,5.

To increase the time interval for comparing the behavior of the radial velocities measured 
from symmetric absorptions V(sym) and H$\alpha$, we added the results of~[14], 
obtained using the same spectrograph and method, to Fig.\,5. Figure~5 shows that V(sym)
and V(H$\alpha$) do not agree for most of the observations. We emphasize that, contrary 
to the claim of Lobel~et\,al.~[24], these velocities turned out to be close or
coincident at several epochs. The velocity differences obtained using observations 
obtained on the same date indicate a dynamic stratification of the upper
atmospheric layers of $\rho$\,Cas.

\begin{figure}[ht]
\includegraphics[angle=0,width=0.8\textwidth,bb=15 45 730 530,clip]{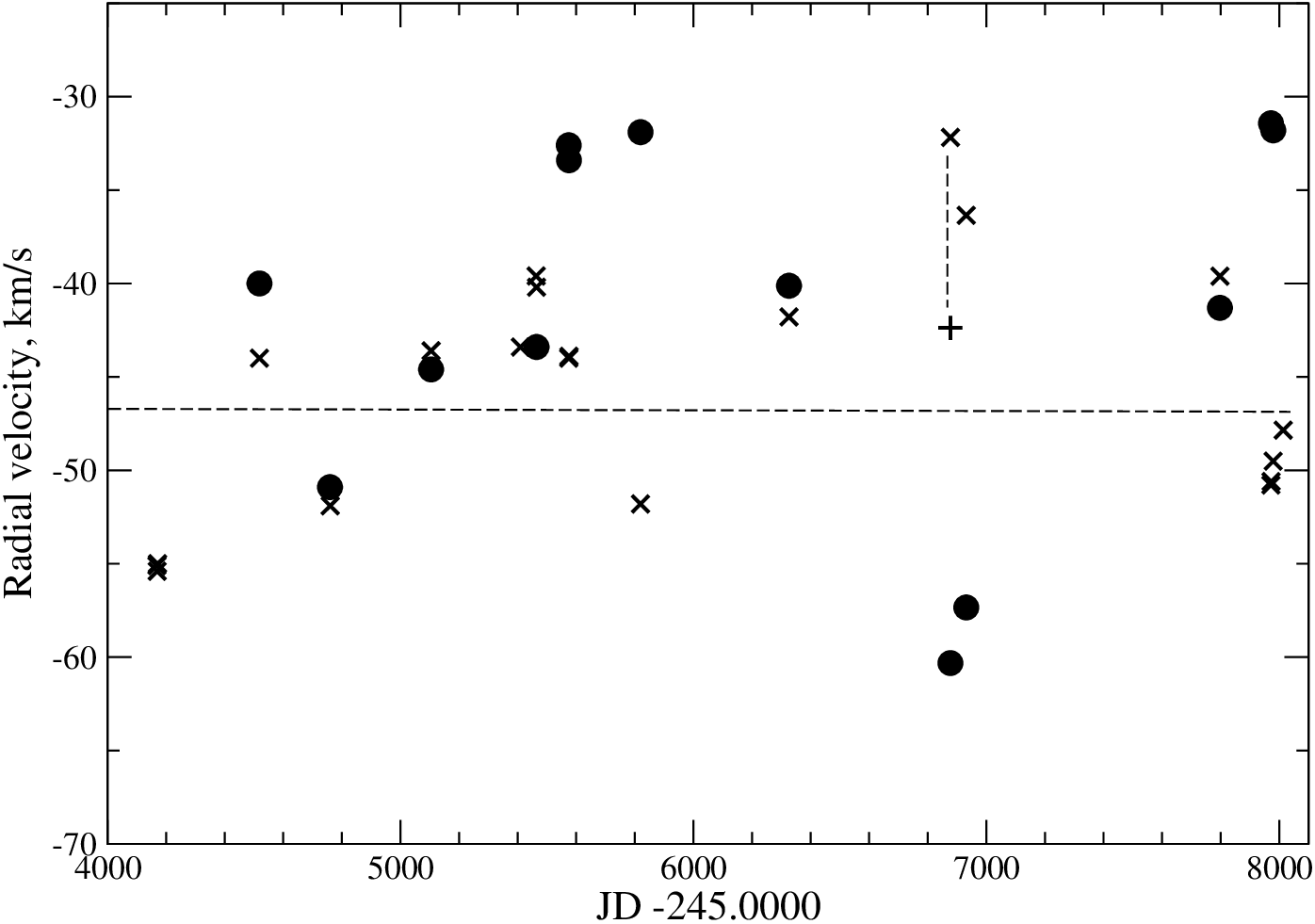} 
\caption{Behavior of the radial velocities measured from symmetric absorption lines (x’s) and the H$\alpha$ 
        line (bold points) in the spectrum of $\rho$\,Cas. Data for epochs preceding JD~2456300 were taken 
        from~[14]. The vertical dashed line connects the velocities measured on August~8, 2014 using symmetric 
        absorptions of ions (plus) and neutral atoms (x).}
\label{Velocity_var}   
\end{figure}

The difference between V(sym) and V(H$\alpha$) is particularly large in August--September~2017, on the descending 
branch of the brightness variations. At the same time, V(sym) and V(H$\alpha$) agree in the spectra taken on 
February~2, 2013 and February~13, 2017, on the rising branch of the brightness variations. However, these 
velocities differ strongly in spectra taken on the rising branch in 2014. It is interesting that the unusual 
H$\alpha$ profile with emission in the line core also first appeared in 2014. It is not by chance that this 
unique profile of splitted  absorptions and additional velocity stratification indicated by lines of ions
and neutral atoms was first detected on August~8, 2014. These changess are obviously due to the fairly
powerful ejection ($\approx 0.5^m$) that occurred from April to November 2013, when the effective temperature of
the star decreased from 7000 to 5000\,K~[15]. As it can be seen from our spectrum for February~2, 2013,
before the beginning of this shell episode, the stellar spectrum and the velocity pattern in the atmosphere
corresponded to the data for 2007--2011~[14]. 

At certain epochs, the velocities measured from absorption lines of ions with high excitation potentials differ 
significantly from those derived from other absorption lines. Such absorption lines include SiII~4028, 4128, 5041, 5055, 
and 5056\,\AA{}, for which the excitation potentials of the lower level is about 10\,eV, SiII~6347 and 6371\,\AA{}, 
for which the excitation  potentials of the lower level is about 8\,eV, and some lines of FeII, for which the excitation 
potentials of the lower level above 5\,eV: FeII~6383, 6443, 6446\,\AA{}. The velocities derived from these high-excitation
absorption lines, which form in the deepest layers of the atmosphere, is always close to the systemic velocity.

The profiles of the NaI~D lines (Fig.\,6) contain broad components, similar to those possessed by the
strongest absorptions, but broader, as well as a narrow absorption feature with Vr\,=$-49.8$\,km/s and
another component close to Vr$\approx  -29$\,km/s that is poorly distinguished in our spectra. The 
absorption  with Vr$\approx -50$\,km/s is formed in the interstellar medium, and indicates a position 
for $\rho$\,Cas beyond the Perseus arm~[28]. The interstellar NaI line corresponding to the Local Arm 
cannot be distinguished in the long-wavelength photospheric component. The features observed in 
the KI~7698\,\AA{} profile shown in Fig.\,6 mainly correspond to those of the NaI~D lines. However, 
interstellar components with Vr$\approx -12$ and $-29$\,km/s, corresponding to the Local
Arm and Perseus Arm, are clearly distinguished in the former. Figure~6 also shows the BaII~6141\,\AA{} 
profile, providing additional support for our conclusion in~[14] that the short-wavelength components of
splitted adsorptions form in the circumstellar medium of the hypergiant.

\begin{figure}[ht!]
\includegraphics[angle=0,width=0.7\textwidth,bb=40 50 720 530,clip]{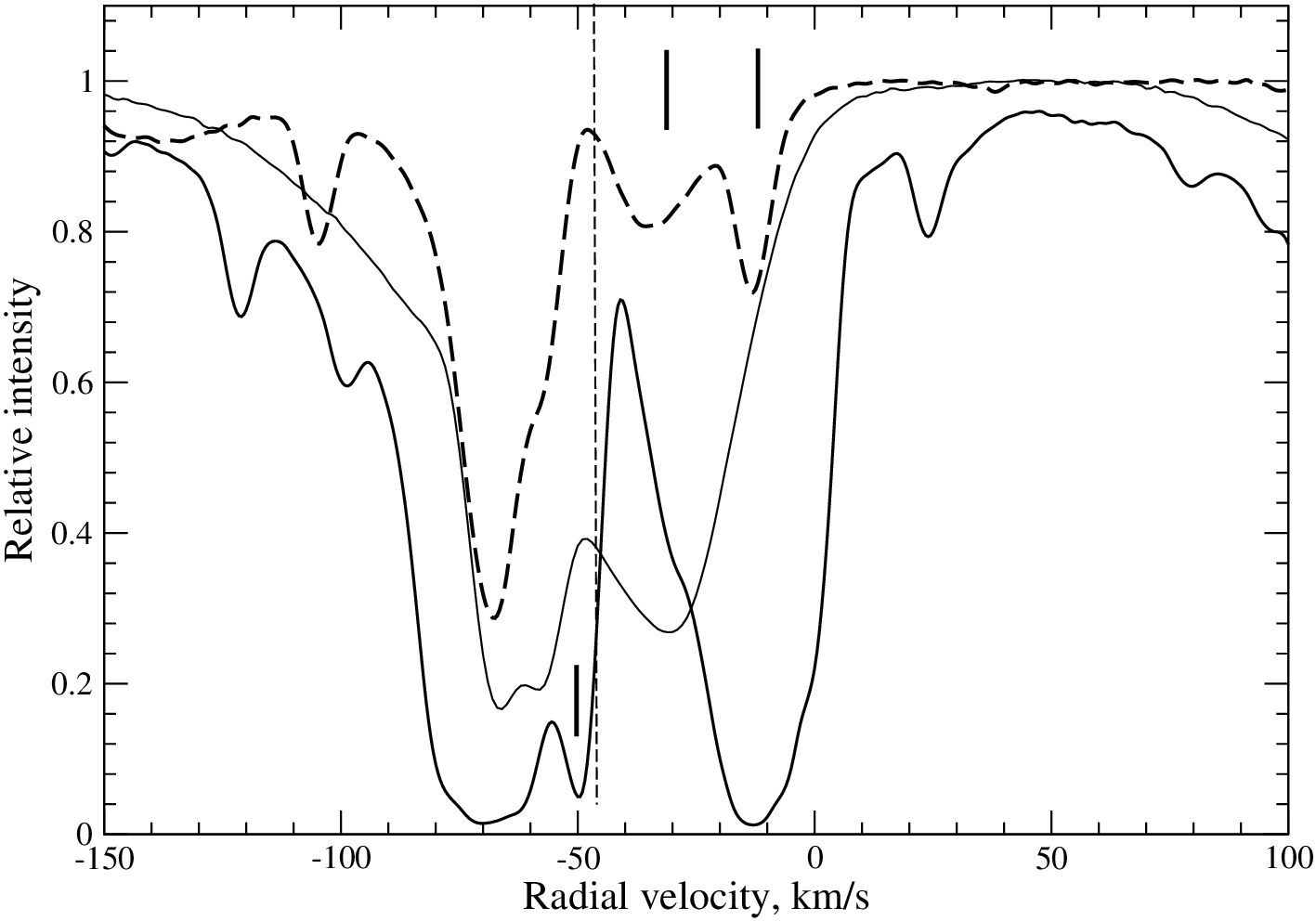} 
\caption{Profiles of selected lines in the spectrum of $\rho$|,Cas obtained on August~6, 2017: D2~NaI~5889\,\AA{} (bold), 
   KI~7698\,\AA{} (dashed), and BaII~6141\,\AA{}  (thin). The vertical bars mark the positions of interstellar lines with 
   velocities V(IS)\,=$-49.7$,  $-29$, and $-12$\,km/s. The vertical dashed line indicates the systemic radial velocity 
   Vsys\,=$-47$\,km/s~[24]. } 
\label{NaK}   
\end{figure}

We should make one more remark about the radial velocities measured using emission and absorption
lines. For absorptions that form close to the photosphere and show low-amplitude pulsations, it is
fair to assume that the radial velocity of the pulsating layers can be obtained by multiplying 
the measured velocity by a projection factor $p$ that depends on the adopted limb-darkening law. 
This factor also differs for various model approximations; for instance, $p=1.41$~[29] and 1.31~[30]. 
Consequently, in the translation from the measured radial velocities to the velocities of the motion 
along the radius, all values derived from absorptions of moderate intensity (about $\pm10$\,km/s) 
should be multiplied by a projection factor $p$. In this case, the amplitude of the pulsation radial 
velocity will increase to $\pm13\div 14$\,km/s. This correction does not apply to emission lines formed in
the outer, diluted  shell, where the same contributions to the profile are made by the near and far 
hemispheres relative to the observer. Figures~2, 3, 4, and 6 compare the observed profiles on the 
scale of the measured radial velocities, and therefore velocities corresponding to the positions 
of the absorption wings of the lines should be increased by a factor of 1.3. In this case,
the pattern of the motions becomes more realistic: outward motions dominate over the systemic velocity. 
However, the emission details of the H$\alpha$ line form under conditions intermediate between 
the outermost comparisand deep layers of the atmosphere. Therefore, strictly speaking, the shape of any complex 
emission-absorption profile should change upon recalculation to velocities of the motion along the radius.
This also applies to the case when the splitting  of strongest absorptions of metals is interpreted as
the result of the appearance of an emission feature in the core of the absorption line~[5], with the position
of this emission feature coinciding with the systemic velocity. If we believe that the emission in lines of
metals arises in a region of emiting behind a shock front, so that we observe emission only from
the near hemisphere, we must explain this surprising coincidence of velocities. In addition, the splitting  of
absorptions into three components that we have found in this study does not fit into a picture with the
formation of a shock front that is at rest relative to the center of the star.

\section{Conclusions}

We have presented the results of an analysis of optical spectra of $\rho$\,Cas obtained shortly before and just 
after its brightness minimum in 2013. Various lines in the spectra of this star, which has an extremely
extended atmosphere, are variable and always show differential shifts, since the regions of their formation
extend from pulsating near-photospheric layers to the non-stationary wind. We have found that the
velocity derived from symmetric absorption lines Vr(sym) is variable, with an amplitude of more
than 10\,km/s higher than the amplitude found in our previous study of $\rho$\,Cas.

Continuation of spectral monitoring using homogeneous material has led to the discovery of new
phenomena. This is also true of the behavior of the H$\alpha$ line. While the H$\alpha$  had an 
almost symmetric absorption profile with the position of the core close to the systemic velocity in 
2008--2011, this profile changed substantially after the shell episode, starting from October~1, 2014, 
when the line acquired a doubled core, with the short-wavelength component shifted by up to 50\,km/s 
relative to the systemic velocity. At different observing dates in 2017, the H$\alpha$ had an inverse 
P\,Cygni  profile (February~13, 2017) and a doubled profile with a large offset toward longer wavelengths 
(August~6, 2017). 
These shifts of the H$\alpha$ proﬁle indicate changes in the structure of the upper layers of the extended atmosphere
and enhanced instability of these layers. The shift of the profile towards longer wavelengths indicates an
infall of the layers where the line is formed. 

A comparison of the radial velocities derived from symmetric absorption lines and from the H$\alpha$ line 
shows that these velocities disagree for most epochs, indicating inhomogeneity of the upper layers of 
the atmosphere and gaseous envelope of the star. 
This difference is particularly large in spectra obtained on the descending branch of the brightness.
The velocities V(sym) and V(H$\alpha$) agree in the spectra obtained on February~2, 2013 and February~13,
2017, when the brightness was increasing. However, V(sym) and V(H$\alpha$) differ considerably in the 2014
spectra, obtained on the rising branch of the brightness variations, when the doubled core of H$\alpha$ 
appeared for the first time. In a spectrum obtained on August~8, 2014, we found additional stratification
of the velocities derived from lines of ions and neutral atoms.

In the long-wavelength region of the spectrum, we identified several weak shell emission lines of iron
group elements in the spectrum taken in 2013; the intensity of these emissions decreased in 2017, with
only a small fraction being visible in the spectrum for September~5, 2017. Shell emission lines of metals
were also detected in the wings of the H and K lines of CaII.

The profiles of strongest absorptions remained splitted in all spectra of $\rho$\,Cas obtained in
2014 and 2017. However, unlike the previous observing seasons, these absorptions have three rather
than two components in the 2017 spectra. This new property of the splitted absorptions supports 
our interpretation of this structure: the short-wavelength components of these lines is formed in the
circumstellar envelope, where weak emissions of metals and one of the broad components of the NaI~D 
doublet are also formed.

The kinematic instability of the upper atmosphere, which is subject to the influence of the stellar wind,
is manifest via the variability of the short-wavelength wing of the BaII~6141\,\AA{} line, which extends to
$-150$\,km/s in the spectrum taken on February~2, 2013, before the mass ejection event. In general,
we observe that the atmosphere of $\rho$\,Cas has experienced the influence of an additional factor during
the ejection episode, which substantially changed the kinematic pattern of its upper layers.

\section*{Acknowledgements}

This study was supported by the Russian Foundation for Basic Research (grants 16--02--00587\,a and 
18--02--00029\,a). We have made use of the SIMBAD, SAO/NASA ADS, and VALD databases.

\end{document}